\documentclass[aip,reprint,showpacs,floatfix,amsmath,amssymb]{revtex4-1}

\usepackage{graphicx, color}
\usepackage{amsmath,amssymb,wasysym}
\usepackage{float}
\usepackage{verbatim}
\usepackage{subfigure}
\usepackage{sidecap}
\usepackage{multirow}
\usepackage{marginnote}
\definecolor{goldenrod}{rgb}{0.85, 0.65, 0.13}

\newcommand{\p}{\textrm{pore}}
\newcommand{\f}{\textrm{force}}
\newcommand{\e}{\textrm{ent}}

\newcommand{\figOne}{
	\begin{SCfigure*}
    \centering
		\includegraphics[width=0.65\textwidth]{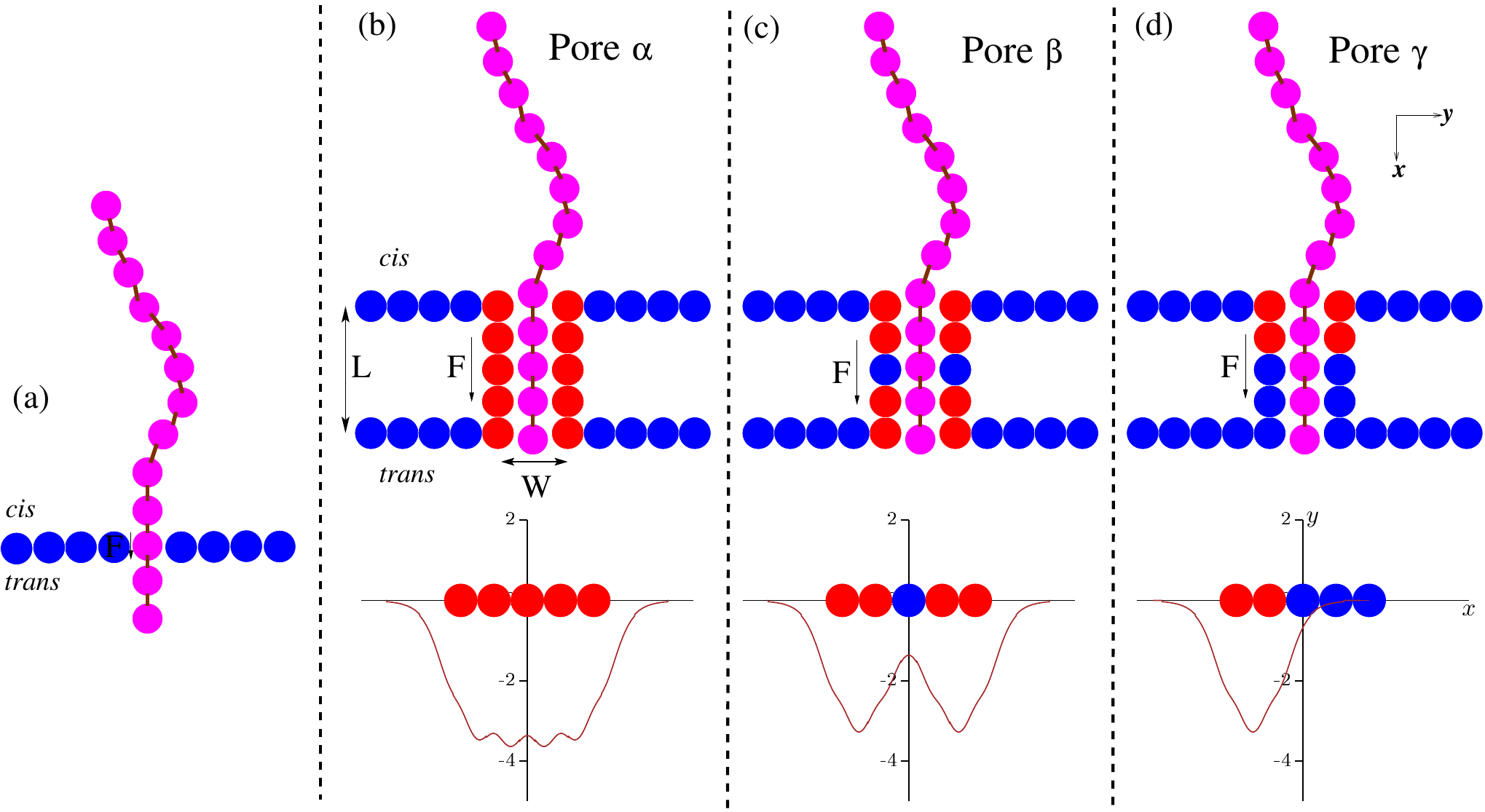}
		
		\caption{\scriptsize Schematic diagram of a semiflexible polymer of stiffness $\lambda$
			translocating from the \textit{cis} to the \textit{trans} side through (a) Pore of
			unit length ($L = \sigma$) and three extended patterned pores (b) Pore $\alpha$ (c)
			Pore $\beta$, and (d) Pore $\gamma$ of length $L$ and width $W$. A driving force
		${\boldsymbol F}_{\textrm{ext}} = F \hat{\boldsymbol x}$ acts on every monomer inside
	the pore. The potential energy of a polymer bead along the pore axis due to pore beads
on either side, for various pore types is also shown. \label{fig:1}}

\end{SCfigure*}
}

\newcommand{\figTwo}{
	\begin{SCfigure*}[][!t]

	\centering
	\includegraphics[width=0.65\textwidth]{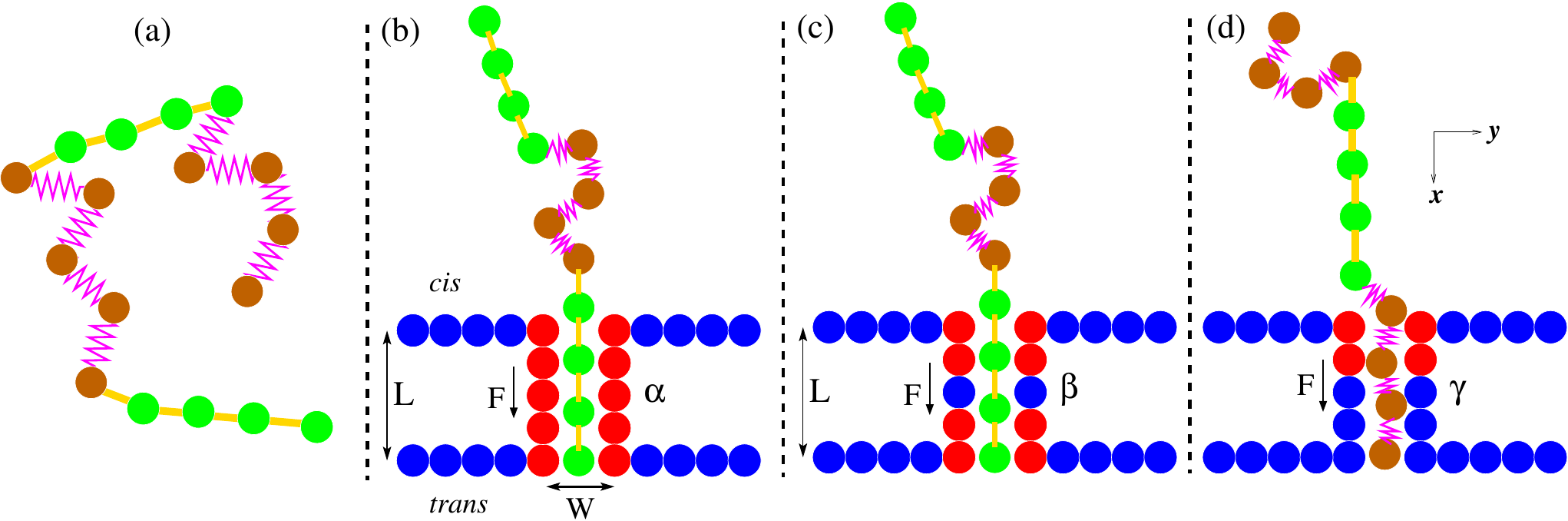}
  
	\caption{\scriptsize (a) Schematic diagram of a polymer with alternate blocks of stiff (S) and
		flexible (F) segments $S_nF_n$ each having $n=4$ bonds. The stiff (S) and flexible (F)
		bonds are shown by straight and zig-zag lines, respectively. (b) and (c) Polymer
		$S_4F_4$ translocating through Pores $\alpha$ and $\beta$, respectively, with the
	stiff end entering the pore first. (d) Polymer $F_4S_4$ translocating trough Pore
$\gamma$ with the flexible end entering the pore first. \label{fig:2}}
 
\end{SCfigure*}
}

\newcommand{\figThree}{
\begin{figure*}[!ht]
	\centering
	\includegraphics[width=\textwidth]{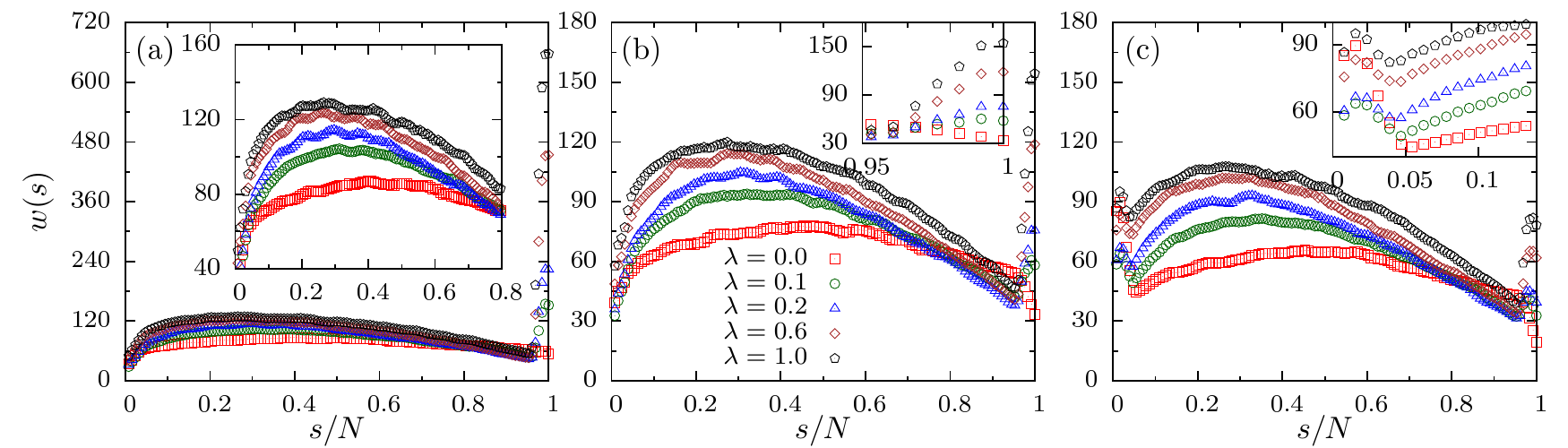}

	\caption{Mean waiting times $w(s)$ for monomers of a semiflexible polymer of various
		stiffness $\lambda$ for (a) Pore $\alpha$, (b) Pore $\beta$, and (c) Pore $\gamma$.
		The inset in (a) represents $w(s)$ for Pore $\alpha$ excluding the end monomers.
		Inset in (b) shows the end monomers region for Pore $\beta$ while inset in (c) shows
	the behavior for the initial monomers entering the pore for Pore $\gamma$.  Note that
error bars are smaller than the point size and are not shown here.} \label{fig:3}

\end{figure*}
}

\newcommand{\figFour}{
	\begin{figure}[b] 
	\centering
	\includegraphics[width=3.0in]{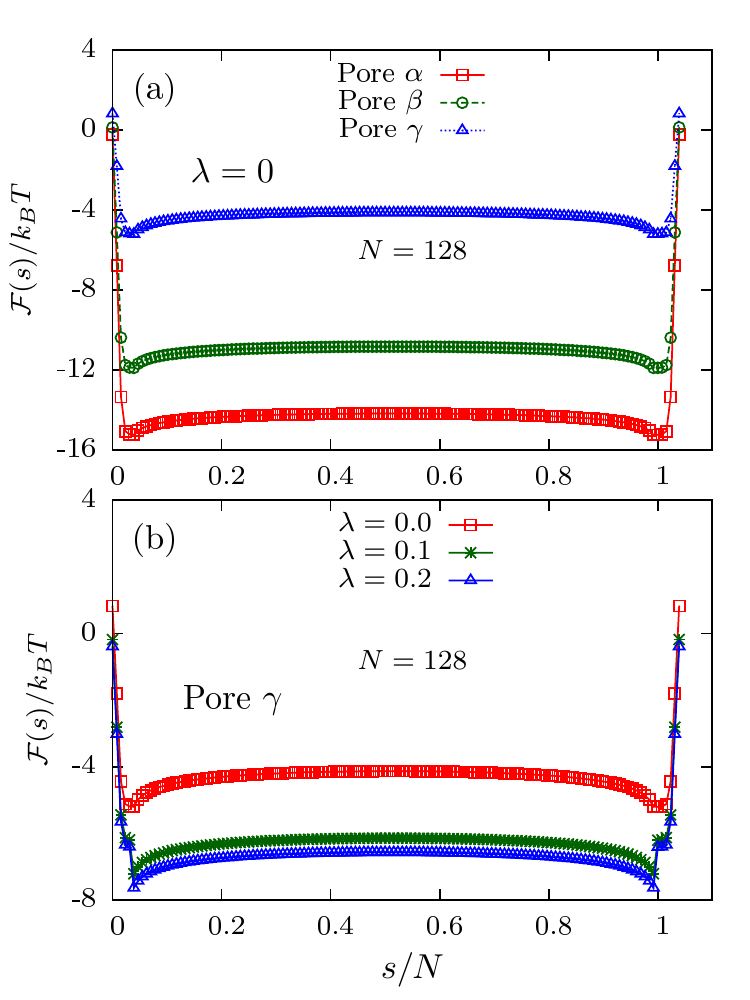}
	
	\caption{(a) Free energy $\mathcal{F}/k_BT$ as a function of $s/N$ for a flexible
		polymer ($\lambda = 0$) of length $N=128$ translocating through various pores.  (b)
		Free energy $\mathcal{F}/k_BT$ as a function of $s/N$ for a polymer of length $N=128$
		translocating through Pore $\gamma$ for various values of chain stiffness $\lambda$.
	In plotting these figures, we have ignored the free energy contribution due to the
external driving force ${\boldsymbol F}_{ext} = F \hat{\boldsymbol x}$. \label{fig:4} }

\end{figure}
}

\newcommand{\figFive}{
	\begin{figure}[b] 
	\centering
	\includegraphics[width=3.0in]{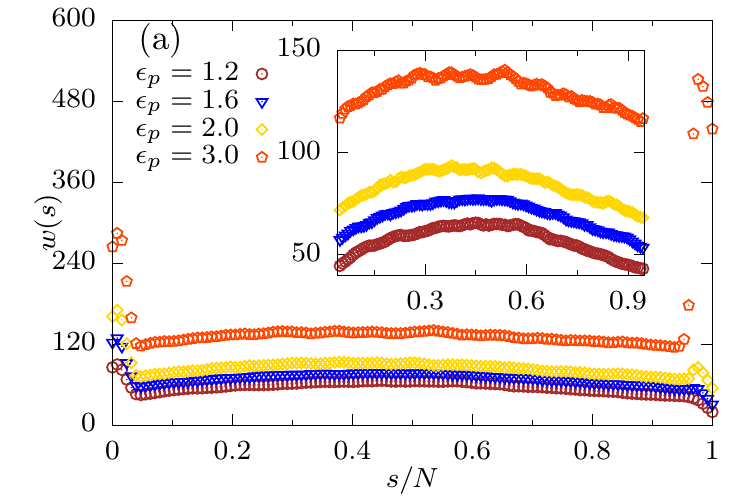}
	\includegraphics[width=3.0in]{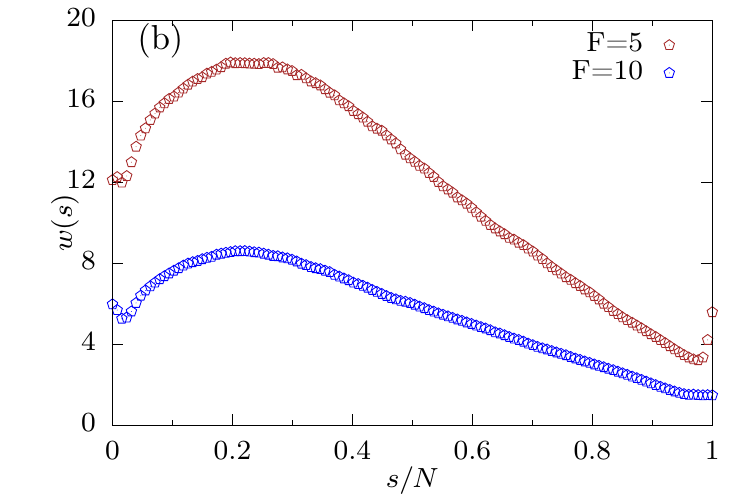}
	
	\caption{(a) The mean waiting times, $w({s})$, as a function of $s/N$ for a flexible
		polymer ($\lambda=0$) for various values of $\epsilon_{\rm pore}$ for Pore $\gamma$ at
		$F=1$. The inset shows the same data for the range $0.1 \le s/N \le 0.9$. (b) The mean
		waiting times, $w({s})$, for higher values of external force $F=5$ and 10 for Pore
	$\alpha$ at $\epsilon_{pore}=1.2$. The sharp rise in $w(s)$ for the end monomers is
completely washed off with increasing driving force. \label{fig:5} }

\end{figure}
}

\newcommand{\figSix}{
\begin{figure*}[t]
	
	\begin{center}
		\includegraphics[width=\textwidth]{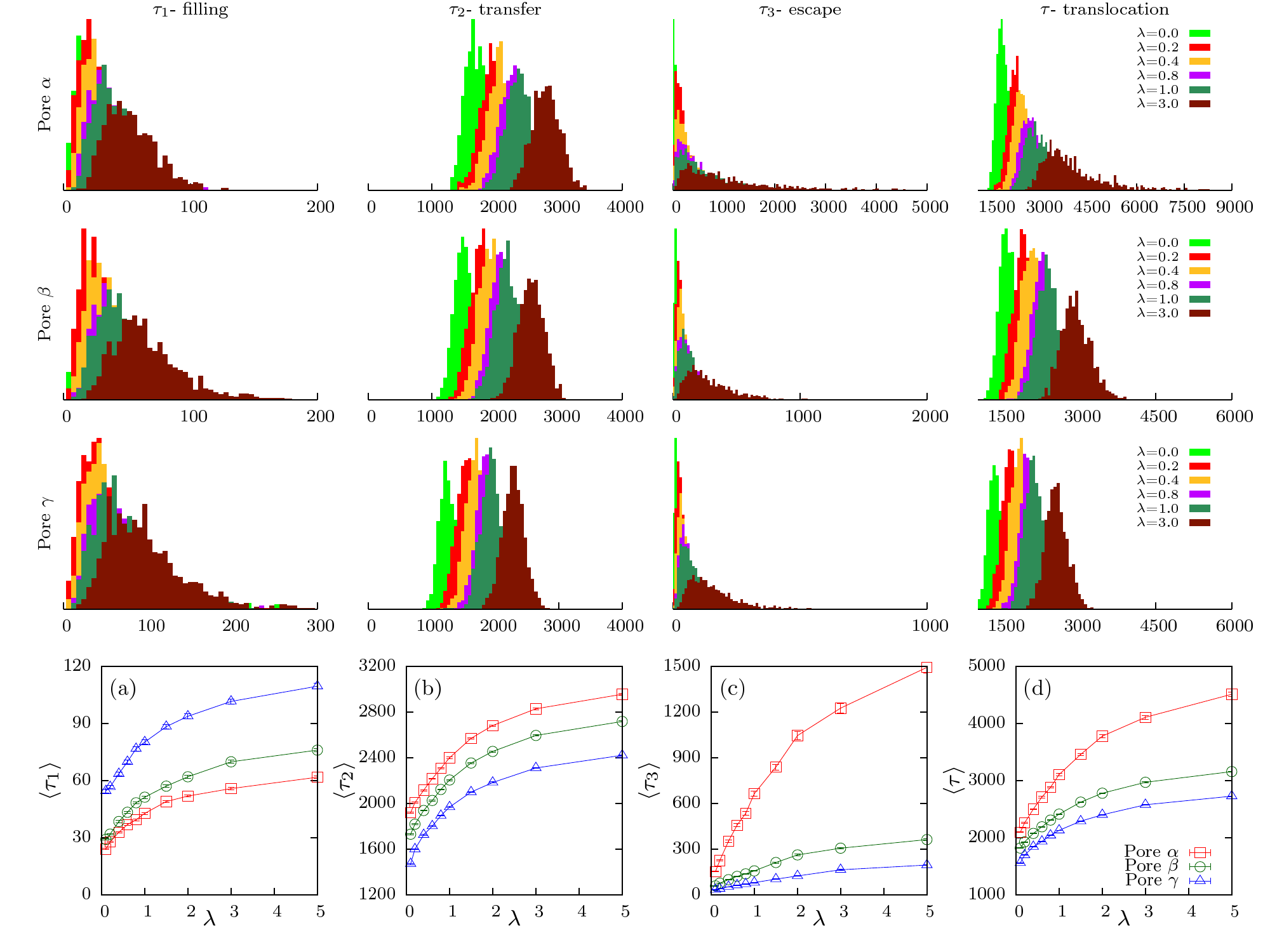}
 	\end{center}

	\caption{Translocation time statistics for semiflexible polymers with homogeneous
		bending rigidity. First, second and third rows : Translocation time distributions for
		pores $\alpha$, $\beta$ and $\gamma$ respectively, for $F = 1.0$, as $\lambda$ is
		varied. Not only are the the three distributions different in their moments across the
		three pores, they also vary with varying $\lambda$.  Fourth row : Average (a) filling
		time, $\langle\tau_1\rangle$ (b) transfer time, $\langle\tau_2\rangle$ , (c) escape
		time, $\langle\tau_3\rangle$ and (d) mean translocation time $\langle\tau\rangle$ as a
	function of $\lambda$. The error bars represent standard deviation of the mean.}
	\label{fig:6} 

\end{figure*} 
}

\newcommand{\figSeven}{
	\begin{figure*}[t]

	\centering
  \includegraphics[width=\textwidth]{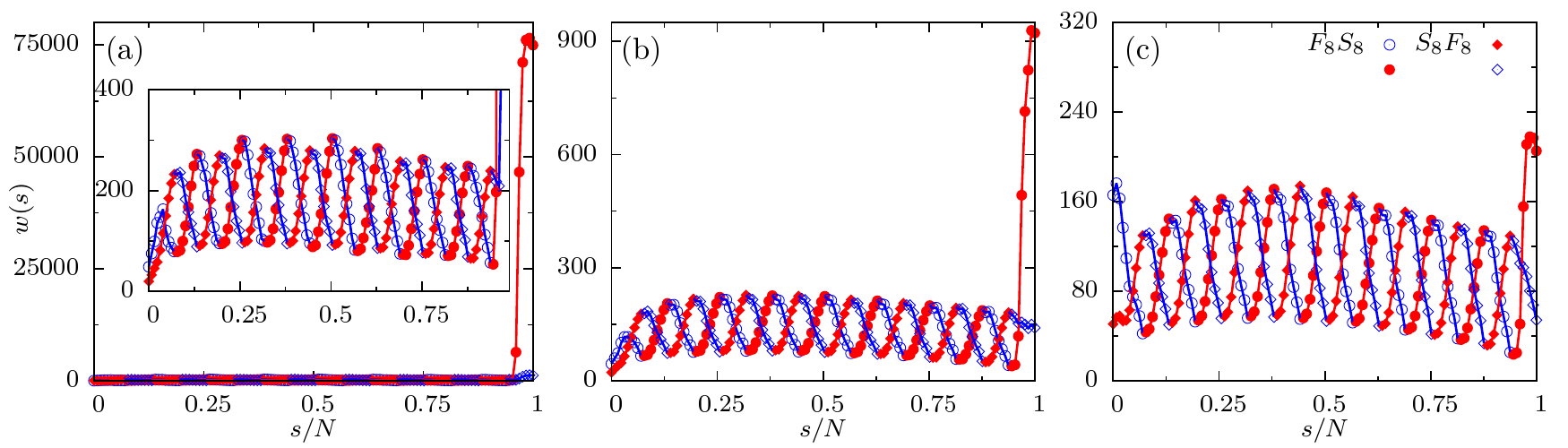}
  
	\caption{ Waiting times for heteropolymers, $S_8F_8$ and $F_8S_8$, translocating through
		(a) Pore $\alpha$, (b) Pore $\beta$, and (c) Pore $\gamma$.  The open and filled
		symbols represent flexible (F) and stiff (S)	segments, respectively. The results for
		heteropolymers, $F_8S_8$ and $S_8F_8$, are shown by circles and diamonds,
	respectively. }\label{fig:7}

\end{figure*}
}

\newcommand{\figEight}{
	\begin{figure}[t]

	\centering
  \includegraphics[width=0.5\textwidth]{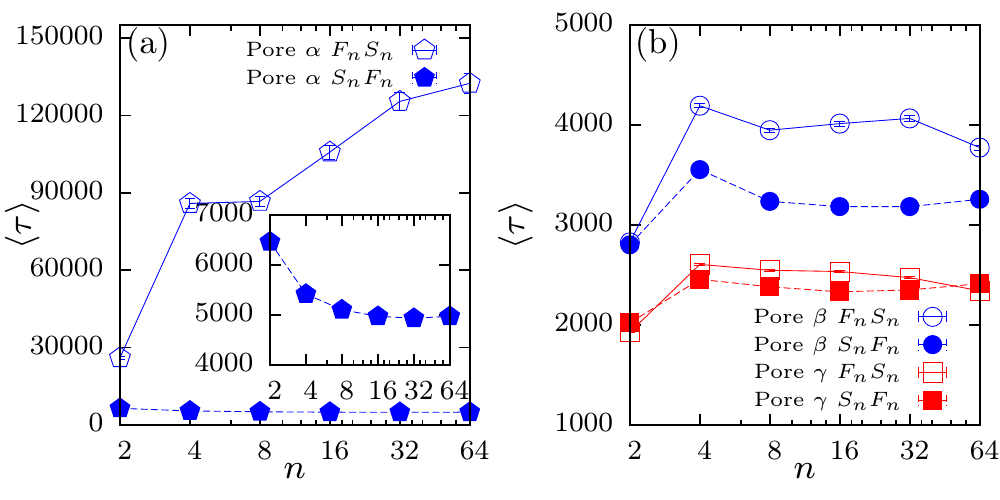}
  
	\caption{Average translocation time $\langle \tau \rangle$ as a function of block length
		$n$ (in log scale) for a polymer of length $N=128$ having alternate stiff (S) and
		flexible (F) segments. (a) Polymer translocating through Pore $\alpha$, and (b) Pores
		$\beta$ (circles) and $\gamma$ (squares). The results for polymer entering the pore
		through the stiff (flexible) end, represented by $S_nF_n$ ($F_nS_n$), are shown by
		filled (open) symbols. The inset in (a) shows the zoomed data for the case where the
		polymer enters the Pore $\alpha$ through the stiff end. The error bars represent
	standard deviation of the mean.} \label{fig:8}
 
\end{figure}
}

\newcommand{\figNine}{
	\begin{figure}[b]

	\centering
  \includegraphics[width=0.45\textwidth]{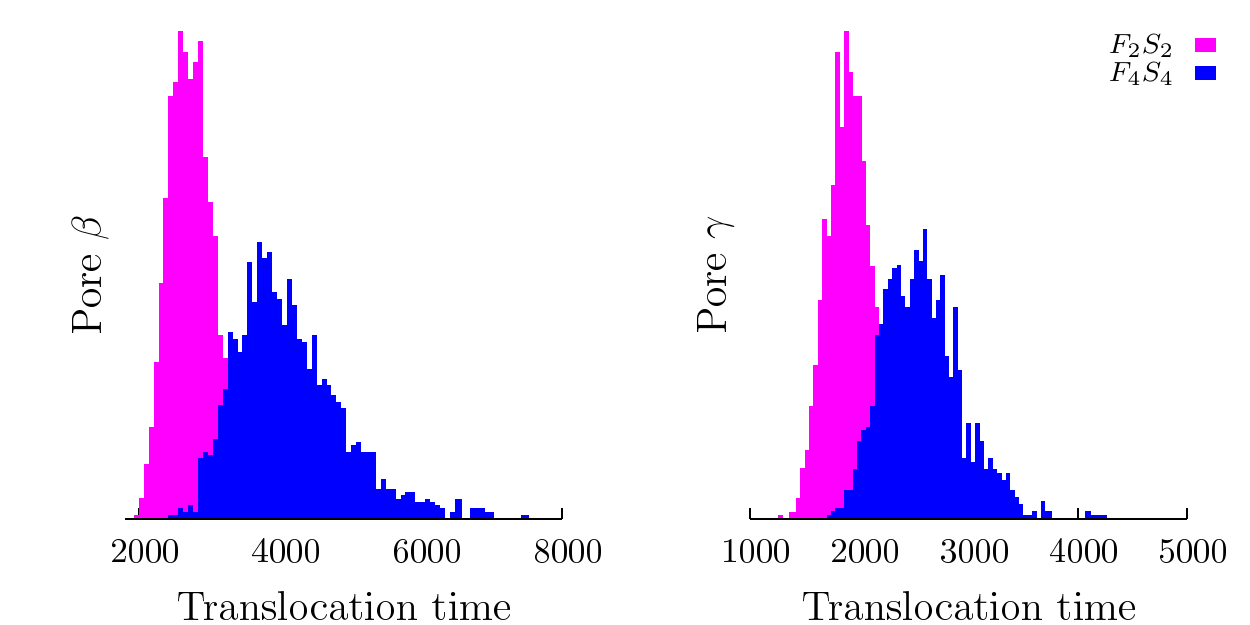}
  
	\caption{Translocation time distribution for sequences $(F_2S_2)_{32}$ and
	$(F_4S_4)_{16}$ translocating through Pores $\beta$ and $\gamma$.} \label{fig:9}
 
\end{figure}
}

\newcommand{\figTen}{
	\begin{SCfigure*}

	\centering
  \includegraphics[width=0.75\textwidth]{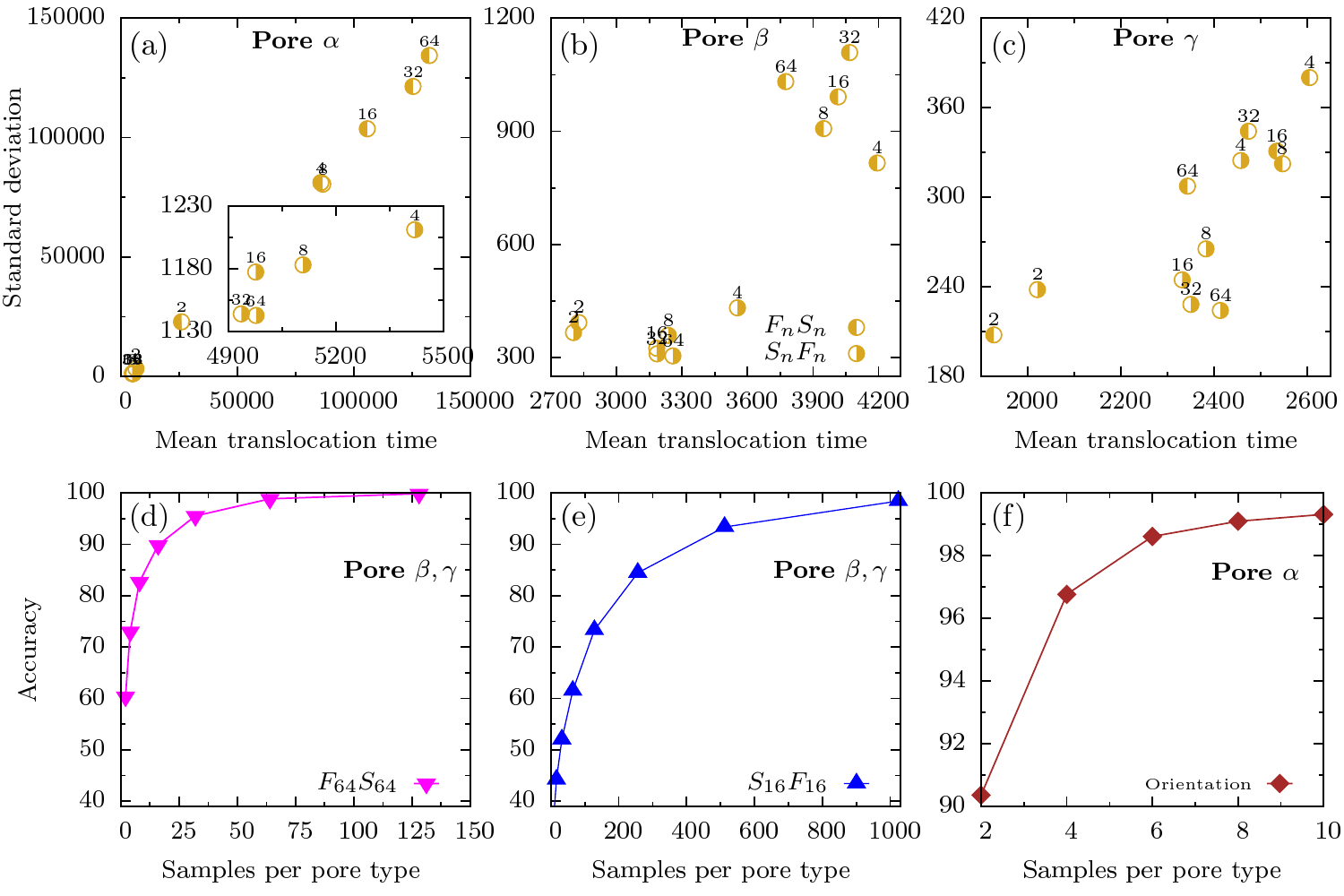}
  
	\caption{\scriptsize [(a) - (c)] Scatter plots showing mean and standard deviation of translocation
		times for Pores $\alpha$, $\beta$ and $\gamma$. The inset in (a) shows the zoomed
		bottom left portion for Pore $\alpha$.  In each of these plots, the polymer entering
		the pore from the stiff end ($S_nF_n$) is shown by symbol half yellow filled circle
		(right), while the polymer entering from the flexible end ($F_nS_n$) is shown by half
		yellow filled circle (left). The accuracy of detecting heteropolymers (d)
		$F_{64}S_{64}$ (pink inverted triangles) and (e) $S_{16}F_{16}$ (blue triangles)
	through Pores $\beta$ and $\gamma$. (f) Accuracy of detection by including Pore $\alpha$
to distinguish orientation of the heteropolymer $S_{16}F_{16}$.}\label{fig:10}

\end{SCfigure*}
}

\definecolor{cream}{RGB}{222,217,201}

\begin{document}

\title{Sequencing of semiflexible polymers of varying bending rigidity using patterned pores}
\author{Rajneesh Kumar}
\email{rajneesh@iisermohali.ac.in}
\author{Abhishek Chaudhuri}
\email{abhishek@iisermohali.ac.in}
\author{Rajeev Kapri}
\email{rkapri@iisermohali.ac.in}

\affiliation{Department of Physical Sciences, Indian Institute of Science Education and
Research Mohali, Sector 81, Knowledge City, S. A. S. Nagar, Manauli PO 140306, India.}

\begin{abstract}

	We study the translocation of a semiflexible polymer through extended pores with
	patterned stickiness, using Langevin dynamics simulations.  We find that the consequence
	of pore patterning on the translocation time dynamics is dramatic and depends strongly
	on the interplay of polymer stiffness and pore-polymer interactions.  For heterogeneous
	polymers with periodically varying stiffness along their lengths, we find that variation
	of the block size of the sequences and the orientation, results in large variations in
	the translocation time distributions. We show how this fact may be utilized to develop
	an effective sequencing strategy. This strategy involving multiple pores with patterned
	surface energetics, can predict heteropolymer sequences having different bending
	rigidity to a high degree of accuracy.

\end{abstract}


\maketitle

\section{Introduction} \label{sec:I}

Polymer translocation is relevant to various biological processes such as the passage of
mRNA through nuclear pores after transcription, horizontal gene transfer in bacterial
conjugation and viral injection of DNA into host cells~\cite{Frank2010,Salman2001}. In the
last two decades, polymer translocation has attracted considerable attention both
experimentally~\cite{Service2006,Lagerqvist2006,Branton2008,Shendure2008,Schloss2008,Persson2010,
Zwolak2008,Min2011,Kasianowicz1996,Braha1997,Akeson1999,Meller2000, Meller2001,Deamer2002,
Meller2003,Maglia2008} and
theoretically~\cite{lubensky,luo2006jcp,muthukumar,Polson2013,matysiak,luo2007prl,luo2008pre,luo2008prl,luo2007jcp,mirigian2012,gauthier,luan,nikoubashman,sung,Muthukumar1999,muthukumar2001,kantor2001,muthukumar2003,metzler,slonkina,kantor2004,milchev2004,gerland,gopinathan,wong,milchev2011,
abdolvahab2011,Cohen2011,Sakaue2007,Sakaue2010,Rowghanian2011,Saito2011,Dubbeldam2012,Ikonen2012,Ikonenjcp2012,Sarabadani2014,Sarabadani2017,Lehtola2009,Haan2010,Bhattacharya2013,Adhikari2013,Adhikari2015,Cohen2012,Cohen2013,Katkar2014}
due to its potential technological applications, such as controlled drug delivery, gene
therapy, and rapid DNA sequencing~\cite{Branton2008}. Experiments have demonstrated that
single stranded DNA and RNA molecules can be electrophoretically driven through biological
and synthetic nanopores~\cite{Kasianowicz1996,Deamer2002}. During polymer translocation,
the ion current flowing through the channel gets blocked, indicating the presence of the
polymer inside the pore. The current blockade readout which could potentially serve as a
signature of the sequence of the polymer segment inside the pore, opens up the
possibilities of efficient sequencing methods~\cite{Kasianowicz1996}. 

Most biopolymers and proteins are semiflexible, implying that the energy cost to bend the
polymer exceeds the entropic propensity to form a random coil
configuration~\cite{Dhar2002,Frey1996}.  The bending stiffness of semiflexible polymers is
quantified by the persistence length, $\ell_{p}$, the length over which the polymer
appears rigid. Experimental studies indicate that sequence dependent bending rigidity is
important for DNA-protein interaction and nucleosome
positioning~\cite{Widom2001,Travers2004}.  Such a dependence is confirmed from cyclization
studies of short DNA fragments, which allows accurate measurement of persistence
length~\cite{Geggier2010}.  Therefore, sequencing techniques based on polymer
translocation, which could correctly sense this variation, necessitates the study of
heteropolymers with varying bending stiffness as they pass through a nanopore. Other
examples of polymers with varying bending rigidity includes partially melted DNA and
proteins which exhibit stiff and flexible segments along the polymer backbone
~\cite{Branden1998,Lee2001,Stellwagen2003,Kowalczyk2010}.

The complex and subtle nature of the translocation process is apparent in the different
theoretical estimates of the exponent $\delta$ reported in the scaling of the mean
translocation time $\langle \tau\rangle$ with the chain length ${\cal L}$, $\langle \tau
\rangle \sim {\cal L}^{\delta}$.  Sung and Park~\cite{sung} and
Muthukumar~\cite{Muthukumar1999,muthukumar2001} considered the translocation process as a
one dimensional barrier crossing problem with the assumption that the translocation time
is long enough to ensure equilibration of the polymer conformations at every stage of the
process. There have been a plethora of later studies predicting different exponents using
arguments like dynamical scaling~\cite{kantor2001,kantor2004}, mass and energy
conservations~\cite{Rowghanian2011} and tension propagation (TP) along the length of the
polymer~\cite{Sakaue2007,Sakaue2010,Saito2011}. TP theory, introduced originally by
Sakaue~\cite{Sakaue2007} for an infinite chain and subsequently modified by Ikonen {\em et
al.}~\cite{Ikonen2012,Ikonenjcp2012} and Dubbeldam {\em et al.}~\cite{Dubbeldam2012} to
finite chains have proved to be successful in explaining the non-equilibrium facets of
driven translocation. In TP theory, the translocation process is described in terms of a
single variable, the monomer index, $s$, at the pore. The part of the translocating
polymer on the \textit{cis}-side is divided into two distinct domains. The external
driving force that acts inside the pore, pulls the monomers nearer to the pore and sets
them in motion. The remaining monomers that are farther away from the pore, do not
experience the pull and on average remain at rest. As the polymer gets sucked inside, more
and more monomers on the \textit{cis} side start responding to the force, with a tension
front separating the two domains propagating along the length of the polymer. At time $t$,
the drag, experienced by a monomer inside the pore, can be written as the sum of the
friction due to the length of the chain in the \textit{cis} side up to which the tension
has propagated, the \textit{trans} side segment and the pore friction.  It is then easy to
see that this drag increases as the tension front propagates and more number of monomers
on the \textit{cis} side get involved.  This increase in the effective friction is
manifested in the mean waiting times, $w(s)$, defined as the amount of time a monomer $s$
spends on average inside the pore. The results from simulation studies of a flexible
homopolymer, passing through a pore of unit length, shows an initial increase with $s$,
implying that the subsequent monomers spend more time inside the pore.  This continues
until the drag becomes maximum when the tension front reaches the last monomer.  The time
at which this happens is called the tension propagation time ($t_{tp}$). At $t_{tp}$, a
maximum number of monomers at the \textit{cis}-side participate in the translocation
process and the monomer $s$, which is inside the pore at that instant, has maximum waiting
time $w(s)$. For $t > t_{tp}$, the system enters the tail retraction stage, where the
monomers on the \textit{cis} side starts decreasing, and therefore the drag decreases, and
so does the waiting time $w(s)$.

The TP theory~\cite{Ikonen2012,Ikonenjcp2012} correctly accounts for the role of pore
friction and thermal fluctuations due to the solvent and their effects on the scaling
exponent.  Further, it explains various values of the exponent observed in previous
studies, thereby providing a unifying picture of polymer translocation. The theory was
recently modified with a constant monomer {\em iso-flux} approximation by Sarabadani {\em
et al.}~\cite{Sarabadani2014, Sarabadani2017}, which leads to a self consistent theory for
polymer translocation with effective pore friction as the only free parameter.
Bhattacharya and others~\cite{Bhattacharya2013,Adhikari2013,Adhikari2015,Haan2010} used TP
theory to explain the dependence of translocation times on the stiffness of a semiflexible
polymer. It was found~\cite{Adhikari2013,Adhikari2015} that the peak of the waiting time
shifts towards lower $s$ indicating that $t_{tp}$ decreases: i.e., the tension propagates
faster along the backbone, as the stiffness of the polymer increases.

A large number of the results discussed above were for pores of small size, where the pore
polymer interactions are negligible.  However, experiments involve pores of finite length,
where pore-polymer interactions play a dominant role. Solid state nanopores with tailored
surface properties~\cite{Storm2003,Kim2006} make it possible to regulate the interactions
of the polymer with the pore as well as reducing
noise~\cite{Ohshiro2006,Iqbal2007,Wanunu2007,Chen2004,Tabard2007}.  Luo {\em et.
al.}~\cite{luo2007prl,luo2008pre,luo2008prl} showed that the mean translocation time of a
polymer across an attractive channel increases with the strength of attraction. This
suggests a possibility to separate polymers with varying interactions with the pore.
Furthermore, translocation dynamics of a heterogenous polymer through an extended pore
show a strong dependence on the sequence. The heterogeneity has been introduced in a
variety of ways.  Luo {\em et. al.}~\cite{luo2007jcp} considered heteropolymers consisting
of two types of monomers which are distinguished by the driving force they experience
inside the pore. The residence time of each bead inside the pore was found to be a strong
function of the sequence. Mirigian {\em et. al.}~\cite{mirigian2012} considered polymers
with differing frictional interaction with the pore and charge. The mean translocation
time of the multiblock polymers depends on the fraction as well as the arrangement of the
blocks.  At a certain optimum length of the charged block, the mean translocation rate is
the slowest. Recent theoretical studies~\cite{Cohen2012,Cohen2013,Katkar2014} considered
channels with varying pore-polymer interactions along its length. The translocation time
distributions showed significant variations across the differently decorated channels.
Katkar and Muthukumar~\cite{Katkar2014} showed that translocation time across a nanopore
of alternate charged and uncharged sections, depends non-monotonically on the length of
the charged section.  In the studies by Cohen {\em et. al.}~\cite{Cohen2012,Cohen2013}, it
was shown that the statistical fluctuations in the translocation time could be utilised
for efficient sequencing of heteropolymers, by suitably engineering pore-polymer
interactions and combining readouts from multiple pores.

\figOne

In this paper, we propose a  sequencing strategy to accurately detect heteropolymers with
sequence dependent bending rigidity through extended patterned pores.  Driving a
homogeneous semiflexible polymer through {\em extended} pores with different patterned
stickiness, we establish the interplay of pore-polymer interactions and polymer rigidity
in determining translocation time statistics. We find that a stiffer polymer takes more
time to translocate through patterned pores, similar to that observed for pores of unit
length~\cite{Adhikari2013}.  However, the mean waiting times of monomers near the pore
entrance and exit vary greatly depending on the pore patterns, a feature distinct for
extended patterned pores.  We utilize these dependencies to test the possibility of
detecting heteropolymers consisting of alternate blocks of stiff and flexible segments, by
passing them through multiple pores. We show that driven translocation of heteropolymers
with varying bending stiffness through extended pores, when coupled to pore patterning,
can lead to efficient sequence detection.

The paper is organized as follows: In Sec.~\ref{sec:model}, we define our model and the
various pore patterns studied in this paper.  In Sec.~\ref{sec:homo}, we discuss results
for the driven translocation of semiflexible polymer of homogeneous stiffness through
extended patterned pores. The results for the driven translocation of semiflexible polymer
consisting of alternate blocks of stiff and flexible segments and the sequencing method
are discussed in Sec.~\ref{sec:hetro}.  Finally, we summarize our results in
Sec.~\ref{sec:summary}.

\section{Model and Simulation Details} \label{sec:model}

\noindent 
{\bf Homopolymer model.} The polymer is modeled as a self-avoiding
semiflexible polymer by using beads and springs in two dimensions (Fig.~\ref{fig:1}).
Semiflexibility is introduced by the bending energy term
\begin{equation}\label{eq:bend}
	U_{\textrm{bend}} = \frac{\kappa_b}{2\sigma} \sum_{i=1}^{N-2}[{\bf t}_{i+1} - {\bf t}_{i}]^2,
\end{equation}
where $\kappa_b$ is the bending rigidity of the polymer, $\sigma$ is the equilibrium bond
length and ${\bf t}_i = [{\bf r}_{i+1} - {\bf r}_i]/b_i$ is the local tangent. Here, $b_i
= |{\bf r}_{i+1} - {\bf r}_i|$ is the instantaneous bond length. $\kappa_b$ represents the
stiffness of the polymer, and in two dimensions it is related to the persistence length as
$\kappa_b/k_BT = \ell_p/2$, where $k_B$ is the Boltzmann's constant and $T$ is the
temperature.

The beads of the polymer experience an excluded volume interaction modeled by the
Weeks-Chandler-Andersen (WCA) potential of the form
\begin{equation}\label{eq:LJmm}
	U_{\textrm{bead}}(r) = \begin{cases}
		4 \epsilon \left[ \left( \frac{\sigma}{r} \right)^{12} - \left( \frac{
		\sigma}{r} \right)^{6} \right] + \epsilon  & \text{for} \ r \le r_{min} \cr
					0  & \text{for} \ r > r_{min},
	\end{cases}
\end{equation}
where, $\epsilon$ is the strength of the potential.  The cut-off distance, $r_{min} =
2^{1/6} \sigma$, is set at the minimum of the potential. Consecutive monomers in the chain
interact via the finite extension nonlinear elastic (FENE) potential of the form
\begin{equation}\label{eq:FENE}
	U_{\textrm{bond}} (r) = - \frac{1}{2} k R^2 \ln \left( 1 - \frac{r^2}{R^2} \right),
\end{equation}
where $k$ is the spring constant and $R$ is the maximum allowed separation between
consecutive monomers of the chain. The length of the polymer is given by $N\sigma$, where
$N$ is the number of beads.

\noindent
{\bf Heteropolymer model.} A heteropolymer is modelled similarly by using beads and
springs with the polymer segment representing $n$ monomers each of stiff ($S$) and
flexible ($F$) beads arranged in symmetric blocks $S_nF_n$. A schematic diagram of such a
polymer with $n=4$ is shown in Fig.~\ref{fig:2}(a). As an example, for a polymer with $N =
128$, the minimum value of $n=1$ is for $(S_1F_1)_{64}$, i.e., 64 repeat units of
$S_1F_1$, and the maximum value of $n = N/2 = 64$ is for a single unit of $S_{64}F_{64}$.
For a heteropolymer, it makes a difference whether a flexible or a stiff end enters the
pore first (Figs. 2(b), 2(c), 2(d)).

\figTwo

\noindent
{\bf Pore model.}
The pore and the wall are made from stationary monomers separated by a distance of
$\sigma$ from each other.  The pore is made up of two rows of monomers symmetric about the
$x$-axis. The length of the pore is taken to be $L$ with a diameter $W$ (see
Fig.~\ref{fig:1}). 

\vskip 0.2cm
We choose extended pores of length $L = 5\sigma$ with three 
different pore patterns :
\begin{enumerate}
	
	\item[(1)] Pore $\alpha$ is an attractive pore. All the monomers of the pore interact
		with the polymer by the LJ potential :
		\begin{equation}\label{eq:LJpm}
			U_{\textrm{pore}}(r) = \begin{cases}
			4 \epsilon_{\textrm{pore}} \left[ \left( \frac{\sigma}{r} \right)^{12} - \left( \frac{
			\sigma}{r} \right)^{6} \right] & \text{for} \ r \le r_{c} \cr
					0  & \text{for} \ r > r_{c},
			\end{cases}
		\end{equation}
		where $\epsilon_{\textrm{pore}}$ denotes the potential depth and $r_c = 2.5 \sigma$ is
		the cut-off distance.

	\item[(2)] Pore $\beta$ has an attractive entrance and exit. The first two and the last
		two monomers of the pore interact with the polymer by the LJ potential, and the middle
		monomer by WCA potential as in the pore of unit length. 

	\item[(3)] Pore $\gamma$ has an attractive entrance and repulsive exit. The first two
		monomers of the pore interact with the polymer by the LJ potential and the last two
		monomers of the pore by WCA potential as above. 
	
\end{enumerate}
The interaction between the wall beads and of the polymer ($U_{\textrm{wall}}$),
is the same as the intramonomer interaction ($U_{\textrm{bead}}$).

To facilitate transfer from the \textit{cis} to the \textit{trans} side of the pore, the
polymer experiences a driving force, ${\boldsymbol F}_{\textrm{ext}} = F \hat{\boldsymbol
x}$ directed along the pore axis with magnitude $F$, which acts on every polymer bead
inside the pore. This mimics the electrophoretic driving of biopolymers through nanopores.
Due to the larger entropic cost involved in confining the polymer in extended pores, the
pore entrance in such cases are chosen to be attractive to initiate the translocation
successfully.  A schematic diagram of semiflexible polymers translocating from the
\textit{cis} to the \textit{trans} side through the pore of unit length and pores
$\alpha$, $\beta$, and $\gamma$ are shown in Fig.~\ref{fig:1}(b)-(d), respectively.

To integrate the equation of motion for the monomers of the chain we use Langevin dynamics
algorithm with velocity Verlet update. The equation of motion for a monomer is given by
\begin{equation} \label{eq:BD}
	m \ddot{\boldsymbol r}_i = - {\boldsymbol \nabla} U_{i} + {\boldsymbol F}_{ext} - \zeta
	{\boldsymbol v}_i + {\boldsymbol \eta}_i,
\end{equation}
where $m$ is the monomer mass,
\[ 
	U_i =  U_{\textrm{bend}} +  U_{\textrm{bond}} + U_{\textrm{bead}} + U_{\textrm{wall}} +
	U_{\textrm{pore}},
\]
is the total potential experienced by a monomer, $\zeta$ is the friction coefficient,
${\boldsymbol v}_i$ is the monomer velocity, and ${\boldsymbol \eta}_i$ is the random
force with mean $\langle \eta (t) \rangle=0$ satisfying the fluctuation-dissipation
theorem $\langle { \eta}_i(t){ \eta}_j(t^{\prime}) \rangle = 2 \zeta k_{B} T \delta_{ij}
\delta(t - t^{\prime})$. 

\figThree

The unit of energy, length and mass are set by the familiar $LJ$ units $\epsilon$,
$\sigma$ and $m$ respectively. This sets the unit of time as $\sqrt{m\sigma^2/\epsilon}$.
In these units, we choose $N = 128$, $L = 5$, $W = 2.25$, $\epsilon_{\textrm{pore}} = 1.2$
(homopolymer), $r_c = 2.5$, $\zeta = 0.7$, $k = 30$, $R = 1.5$ and $k_BT = 1.2$, in our
simulations. These parameters are in accordance with earlier Langevin dynamics simulations
for polymer translocation~\cite{luo2007prl,luo2008prl,Cohen2011,Cohen2012,Cohen2013}.  The
choice of pore width ensures single-file translocation of the polymer and avoids the
formation of hairpin configurations inside the pore.  The stiffness of the semiflexible
polymer is characterized by the dimensionless parameter $\lambda = \ell_p/\ell$ ($\ell$
being the average contour length of the polymer). For the heteropolymer, we choose
$\epsilon_{\textrm{pore}} = 2$, $F=1$ and $\lambda = 0.5$ for the stiff segments. The
choices of $F$,  $\epsilon_{\textrm{pore}}$ and $\lambda$ will be discussed in
Secs.~\ref{sec:IIIA} and \ref{sec:hetro}. A time step of $\Delta t = 0.001$ is used in all
simulation runs.

To initiate the translocation process the polymer has to find the pore.  We start with a
chain configuration with the first bead placed at the entrance of the pore. In order to
get equilibrium initial conditions, we fix the first bead while the remaining beads of the
chain are allowed to fluctuate.  The first bead is then released and the translocation of
the polymer across the pore is monitored.

The translocation time $\tau$ is defined as the time elapsed between the entrance of the
first bead of the polymer and the exit of all the beads from the channel.  All failed
translocation events are discarded. The maximum run time of our simulation is
$5\times10^8$ steps. To calculate statistical properties, we have considered $1500-2000$
successful translocation events.

\section{Translocation of Homogeneous semiflexible polymer} \label{sec:homo}

\subsection{Mean waiting time for extended patterned pores} \label{sec:IIIA}

We provide a qualitative description of the effects of pore patterning on the mean waiting
times and the translocation time distributions, based on the surface energetics of the
pores.  Note that the effects of pore-polymer interactions on polymer translocation has
been extensively studied in the past~\cite{luo2007prl,luo2008prl,Cohen2011,Cohen2012}.
Our study looks at the combined effects of chain flexibility and pore-polymer interactions
on the translocation dynamics.

For the extended patterned pores in our simulations, we calculate the mean waiting time of
a monomers as it translocates from the \textit{cis} to the \textit{trans} side. The
waiting time of a monomer for the extended pore is obtained by calculating the time spent
by it inside the pore, from its entry at the \textit{cis} side to its exit from the pore
at the \textit{trans} side. We observe that the gross features, like a peak in the waiting
times, and the dependence on chain stiffness, as observed earlier~\cite{Adhikari2013} for
the polymer translocation through pore of unit length, are reproduced. Further, we found
additional features near $s = 1$ and $s = N$, which can be attributed to the pore polymer
interactions. More specifically, we note that for Pore $\alpha$, the waiting time $w({s})$
shows a sharp rise in the large $s$ limit (Fig.~\ref{fig:3}(a)). This feature persists for
Pores $\beta$ and $\gamma$ as well, although it is less pronounced. Pore $\gamma$ shows an
initial dip in $w({s})$(Fig.~\ref{fig:3}(c)). For monomers in the bulk of the polymer, the
non-monotonic variation of $w({s})$ as predicted from TP theory for pores of unit length
persists.

\figFour

\figFive

\figSix

\figSeven

In order to understand these features, we focus on the surface energetics of the various
patterned pores. A comprehensive picture of the translocation process, which takes into
account the pore-polymer interactions and entropic contributions, can be obtained by
constructing a free energy landscape, $\mathcal{F}/k_BT$, in terms of the translocation
coordinate $s$\cite{muthukumar2003,Katkar2014}.  The translocation process is separated
into three stages: (i) the pore \textit{filling}, (ii) the \textit{transfer}, and (iii)
\textit{escape} from the pore. At every stage, the free energy of the system has
contributions from (i) pore-polymer interactions, ${\mathcal F}_{\p}$, (ii) polymer
entropy, ${\mathcal F}_{\e}$, and (iii) energy due to the externally applied force,
${\mathcal F}_{\f}$.  In this analysis, we have neglected the contribution to the free
energy due to the constant external force acting on every bead inside the pore, ${\mathcal
F}_{\f}$. The presence of this external force facilitates entry and exit of the polymer
and is therefore expected to influence pore \textit{filling} and \textit{escape} stages.
However, in this study, we restrict ourselves to small forces, where the effects of pore
polymer interactions and polymer entropy are dominant.  The free energy contribution due
to pore-polymer interaction, $\mathcal{F}_{\p}(s)$, is obtained by summing over the LJ
potential (Eq.~\ref{eq:LJpm}) felt by each polymer bead (inside the pore) due to the pore
beads. The entropic contribution for a chain with $s$ monomers on the \textit{cis}(or
\textit{trans}) side is given by the entropy for a polymer with one end fixed to a
wall\cite{muthukumar2003}, $\mathcal{F}_{\e}(s) = k_{B}T (1 - \Gamma) \ln (s)$, where
$\Gamma = 0.69$. 

In the pore filling stage ($0 < s < L$), $s$ monomers are inside the pore and remaining
$N-s$ monomers are in the \textit{cis} side. Then, $\mathcal{F}(s) =  \mathcal{F}_{\p}(s)
+ \mathcal{F}_{\e}(N - s).$ In the transfer stage ($L < s < N$), $L$ monomers are inside
the pore, $s - L$ are on the \textit{cis} side and $N - s$ are on the \textit{trans} side.
Therefore, the free energy at this stage, $\mathcal{F}(s) = \mathcal{F}_{\p}(L) +
\mathcal{F}_{\e}(N - s) + \mathcal{F}_{\e}(s - L).$ In the final stage $(N < s < N + L)$,
$s - L$ monomers are on the \textit{trans} side while $N + L - s$ monomers are inside the
pore.  At this stage, $\mathcal{F}(s) = \mathcal{F}_{\p}(N + L - s) + \mathcal{F}_{\e}(s -
L)$.  In Fig.~\ref{fig:4}(a), we have plotted the free energy $\mathcal{F}(s)/k_B T$ for a
flexible polymer ($\lambda = 0$) as a function of the translocation coordinate for
translocation from Pores $\alpha$, $\beta$ and $\gamma$. 

From Fig.~\ref{fig:4}(a) and the plot of the potential energy experienced by a chain
monomer at any point along the axis of the pore due to the pore beads on either side
(Fig.~\ref{fig:1}), it is clear that Pore $\alpha$ is an uniformly attractive pore.
Although this makes it easier to pull the polymer inside the pore, the attractive
interaction makes it difficult to exit the pore from the \textit{trans} side for small
external forces. Pore $\beta$ has a shallower well compared to Pore $\alpha$.  Further,
the free energy barrier at the \textit{trans} end of the pore is significantly larger for
Pore $\alpha$ than Pores $\beta$. Therefore, the mean waiting times for the end monomers
are significantly less for Pore $\beta$ as compared to Pore $\alpha$.  For Pore $\gamma$,
which has a repulsive exit, this effect is the least.  However, Pore $\gamma$ has a large
repulsive exit and a short attractive entrance. Since the attractive entrance spreads over
two monomers of the pore, only a few beads will be sucked inside the pore initially. Once
it becomes energetically favorable for the first monomer to exit the pore by crossing the
barrier, the inter chain interaction ensures that the monomers immediately adjacent to it
are dragged out resulting in smaller waiting times successively for the initial beads.
This is the cause of the dip in the waiting times for the first few monomers.

The free energy analysis can be extended for the case of a semiflexible polymer ($\lambda
\ne 0$) by replacing the number of monomers $n$ in the entropic contributions with the
number of Kuhn segments $n^{\prime} = n \sigma/\ell_K$, where $\ell_K = 2 \ell_p$
represents the Kuhn length~\cite{gauthier}. Within this approximate approach, we can
explain qualitatively the observed dependencies of the translocation times on the
stiffness $\lambda$ of the polymer. As observed in Fig.~\ref{fig:4}(b),
$\mathcal{F}(s)/k_B T$ shows an increasing well depth as $\lambda$ is increased. This
indicates that it become increasingly difficult to cross the free energy barrier at the
\textit{trans} end of the pore, resulting in increased translocation times with increasing
stiffness.

In our simulations for extended pores, we choose $\epsilon_{pore} = 1.2$ and $F=1$ to
ensure that the effect of pore-polymer interactions, and hence the pore patterning, are
dominant.  Higher $\epsilon_{pore}$ increases the mean translocation time, while at high
values of force, pore patterning effects are significantly reduced. In Fig.~\ref{fig:5}(a)
we see that the mean waiting time for a monomer inside the pore increases with increasing
$\epsilon_{\textrm{pore}}$.  Further, the characteristic features near the \textit {cis}
and \textit {trans} ends become even more prominent vindicating our earlier arguments.
Increasing $F$, the mean waiting time drops sharply and the end features are completely
washed out (Fig.~\ref{fig:5}(b)).

\subsection {Translocation time distributions} \label{sec:IIIB}

In the light of the above free energy argument, it is useful to divide the total
translocation time as $\tau = \tau_1 + \tau_2 + \tau_3$~\cite{luo2007prl,Cohen2012} where
(i) $\tau_1$ is the initial \emph{filling} time, the time taken by the first monomer of
the polymer to reach the exit without returning to the pore, (ii) $\tau_2$, the transfer
time, the time taken from the exit of the first monomer into the \emph{trans}-side to the
entry of the last monomer from the \emph{cis}-side and (iii) $\tau_3$, the escape time,
the time between the entry of the last monomer in the pore and its escape to the
\emph{trans}-side.  We compare the separate average time scales for filling, transfer and
escape for the three different pore patterns to investigate the effect of changing
pore-polymer interactions (Fig.~\ref{fig:6}).

\noindent \emph{Effect of stiffness.}  All time scales show a monotonic increase with
increasing stiffness. This behavior is expected from the discussion of waiting times which
increases with increasing $\lambda$. Note that, unlike a pore of unit length, the
summation of mean waiting times for all monomers do not give the mean translocation time
for an extended pore.  However, when scaled by the pore length, the sum of mean waiting
times still provide an excellent measure of the mean translocation time for such cases
(data not shown).

\noindent {\em Effect of pore patterning.} Our simulation shows that $\langle \tau_1
\rangle$ is the minimum for Pore $\alpha$ and maximum for Pore $\gamma$ for a fixed value
of stiffness $\lambda$. From the free energy diagram, we note that for the initial filling
process ($0<s<L$), the free energy falls sharpest for Pore $\alpha$.  This indicates that
filling is considerably easier for Pore $\alpha$ and less so for Pores $\beta$ and
$\gamma$ which explains the simulation results. The free energy diagram also tells us that
both transfer and escape are dictated by shallowness of the free energy landscape and the
barrier experienced during the expulsion process of the polymer, both of which are maximum
for Pore $\alpha$ and minimum for Pore $\gamma$. This is consistent with the observation
of $\langle \tau_2 \rangle$ and  $\langle \tau_3 \rangle$ for the three different pores at
a given stiffness.

These results can be compared with those earlier observed for flexible
chains~\cite{Cohen2012} as a function of pore-polymer interaction strength,
$\epsilon_{\textrm{pore}}$. As one would expect, the effect of varying
$\epsilon_{\textrm{pore}}$ is quite drastic and was used to demonstrate
sequencing~\cite{Cohen2012} based on its variation. Our analysis on the other hand,
clearly demonstrates the significant effect of chain stiffness on translocation time
distributions for patterned pores without changing $\epsilon_{\textrm{pore}}$.  

\section{Heterogeneous translocation.} \label{sec:hetro}

We next investigate the possibility of heteropolymer sequencing by passing them through
multiple patterned pores. As elaborated in Sec~\ref{sec:model}, we introduce heterogeneity
by varying the stiffness of the polymer along the chain backbone. The heterogeneity
introduced in our polymer model is periodic with alternate flexible ($\lambda = 0$) and
stiff ($\lambda \neq 0$) segments. Note that heteropolymer sequencing have been studied in
the past using flexible polymers where the heterogeneity was introduced in a manner in
which alternate polymer segments interacted with the pore~\cite{Cohen2012}. In our
analysis, we study the experimentally relevant scenario of varying bending rigidity in
biopolymers. Due to this heterogeneity in stiffness, it is important to understand the
effect of switching orientation of the polymer as it translocates from the \textit{cis} to
the \textit{trans} side. We first discuss the effect of heterogeneity and orientation on
mean waiting time and translocation time dynamics.  Our choice of
$\epsilon_{\textrm{pore}} = 2$ and $\lambda = 0.5$ for the stiff segment ensures a
significant difference in the translocation times of the flexible and the stiff segments
across different pores.

\subsection{Waiting times}
\figEight

The effect of tension propagation in a polymer with periodically varying bending rigidity
becomes clear when we look at the waiting time distribution.  The distinct difference in
the behavior of mean waiting times observed in Fig.~\ref{fig:7}, as opposed to earlier
studies using heteropolymers of alternate stiff and flexible segments translocating
through pores of unit length~\cite{Haan2010,Adhikari2015}, are in the edge monomers where
the pore synergetics becomes dominant. For the rest of the monomers, the oscillatory
behavior displays the same characteristics. From waiting time distribution of monomers of
a homogeneous polymer we know that (i) tension propagates faster for chains with
increasing stiffness and hence (ii) leads to larger waiting times. In the case of
heterogeneous polymers, tension propagates intermittently through blocks of stiff and
flexible segments leading to the oscillations in the waiting time
distribution~\cite{Haan2010,Adhikari2015}.  A stiff block has a larger waiting time,
followed by a flexible block with lower waiting time and so on. When the orientation of
the chain is reversed, the oscillations for $S_nF_n$ and $F_nS_n$ are exactly out of phase
as expected. The waiting times for the end monomers of the chain however show distinct
features for different orientations of the chain.
 
In sync with its homopolymer counterpart, the end chain dynamics of heteropolymers is
strongly influenced by the pore-polymer interactions. For Pores $\alpha$ and $\beta$, the
attractive interactions near the \textit{trans} side of the pore dominate, leading to
large waiting times.  Evidently, the waiting times for the end monomers of the chain are
significantly larger for Pore $\alpha$ compared to Pore $\beta$. This effect is
significantly less for Pore $\gamma$ which has a repulsive exit. 

\figNine

\figTen

The end chain dynamics for the reversed conformation $S_nF_n$ is not significantly
affected by these interactions. Pore $\alpha$ due to the large potential barrier does make
it difficult for the end monomers to exit the pore leading to larger waiting times.
However, the waiting times are considerably less compared to $F_nS_n$. Pore $\beta$ and
$\gamma$ are largely unaffected. This is expected from our earlier analysis of larger
waiting times for stiffer chains. Polymer in the conformation $S_nF_n$ enters the pore
with the stiffer block entering first followed by a flexible block. This implies that a
flexible block exits the pore last in this conformation. In contrast, in the conformation
$F_nS_n$, it is a stiff block which exits the pore last from the \textit{trans} side in
the translocation process leading to much larger waiting times. 

We argue that these distinguishing features observed for the end monomers and dominated by
the pore synergetics, result in distinct translocation time distributions for different
pores. We use this to effectively distinguish heteropolymers with varying bending rigidity
using a statistical analysis based on the moments of the distributions. 

\subsection{Average translocation time}

The mean translocation times for the heteropolymers as they pass through the patterned
pores, mimic the behavior of the mean waiting times of individual monomers. For pore
$\alpha$, the difference in the mean waiting times for the two different orientations is
significant and increases with increasing block length $n$ (Fig.~\ref{fig:8}). This is a
result of the difference in waiting times of the end monomers during exit. Also note that
for longer block lengths the tension can propagate over larger lengths of the polymer
uninterrupted. For pore $\beta$, the effect of the longer waiting times for end monomers on
the total translocation time is less significant while for pore $\gamma$, it is effectively
the same for both orientations of the polymer during translocation.

\subsection{Sequencing of polynulceotide with varying bending rigidity}

The sensitivity of the translocation dynamics on the varying bending rigidity of
heteropolymers and the patterning of pores opens up the possibility of sequencing
heteropolymers based on their unique translocation time statistical properties.  For
example, in Fig.~\ref{fig:9} we show the translocation time distribution for sequences
$(F_2S_2)_{32}$ and $(F_4S_4)_{16}$ for pores $\beta$ and $\gamma$. These translocation
time distributions exhibit distinct features corresponding to the variation in the block
lengths of the heteropolymer. However, for certain heteropolymers, there could be
significant overlap in the distribution.  We calculate the mean translocation time
($\langle \tau \rangle$) and the standard deviation ($\sqrt{\langle \tau^2\rangle -
\langle \tau\rangle^2}$) of the translocation times from these distributions and construct
scatter plots for each pore type as shown in Fig.~\ref{fig:10}. The scatter plots reveal
several interesting features. For example, Pore $\gamma$ cannot distinguish between
$(S_{32}F_{32})_2$ and $(S_{64}F_{64})_1$, but Pore $\beta$ can. Similarly, Pore $\beta$
cannot distinguish between $(S_2F_2)_{32}$ and $(F_2S_2)_{32}$, but Pore $\gamma$ can.
These differences in the scatter plots for the various pores, marking the mean and
standard deviation for each sequence, clearly shows that a combination of translocation
time measurements from multiple pores could be utilized to differentiate and identify
sequences with a relatively small number of samples per pore type, which would otherwise
be difficult to distinguish using measurements from a single pore.

Our simulation methodology for sequence detection is as follows. We choose a sequence of
the heteropolymer (say $S_kF_k$, with a specific orientation) from the set of all
available sequences (defined as the training set) used to plot Figs.~\ref{fig:10}(a-c) and
call it an ``unknown'' sequence. This sequence is then passed through a single pore of
$\alpha$, $\beta$ or $\gamma$ type. For each pore type, the heteropolymer is passed
multiple times and successful translocation events are registered (say $m$). For every
attempted translocation, the chain configuration of the heteropolymer is chosen from the
equilibrated configurations obtained in accordance with the simulation strategy discussed
before (Sec.~\ref{sec:model}).  Having registered the successful translocation events
across each pore, the mean translocation time and standard deviation are calculated for
each pore type, which correspond to respective points in the scatter plots. These numbers
are then compared with those of the training set for the corresponding pore using a
``distance'' metric.  The larger the distance between the point corresponding to the
``unknown'' sequence is from a particular known sequence in the scatter plot, the greater
is the relative error for that sequence.  The total error, which is the sum of the
distance from a particular known sequence in all the plots, is minimized to predict the
``unknown'' sequence.  If the predicted sequence matches the sequence we started with,
then this marks a successful detection.

The ratio of the number of times a sequence is correctly detected to the total number of
attempts, gives the accuracy of the measurement process (see Fig.~\ref{fig:10}(d)-(f)).
The samples per pore type merely indicate the number of registered successful
translocation events chosen for the unknown sequence across each pore type. Evidently, if
we use a very large number of samples for a given pore type, the sequence detection would
be accurate. However, this scheme suggests that a combination of different pore types
gives very high accuracy of prediction for a relatively small number of copies of each
pore. In Fig.~\ref{fig:10}, we have used the statistical data for only two pore types,
Pores $\beta$ and $\gamma$, to test our hypothesis. Employing the above scheme we found,
for example, that the accuracy of detection for the sequence, $F_{64}S_{64}$
(Fig.~\ref{fig:10}(d)), reaches $100\%$, even for $~\sim 130$ samples per pore type. It is
important to note that in our method of sequence detection, we have used only the first
two moments of the probability distribution of translocation times. As observed from the
results of the distributions, this is far from accurate. Inclusion of higher moments would
most definitely improve the accuracy of the scheme and lead to a more rapid convergence.
Further, our method has considered only a few possible pore types. It would be interesting
to design pores leading to even more distinguishable translocation time statistics, which
when used in conjunction with varying semiflexibility across the polymer backbone, would
lead to enhanced sequence detection.

From the set of pores chosen for this study, it is evident that for Pore $\alpha$, there
is an order of magnitude difference in the translocation times of stiff and flexible
segments. Therefore, this pore is an ideal candidate to detect the difference in
orientation between $S_kF_k$ and $F_kS_k$ and make the detection process even more
precise. Indeed, we find that the accuracy of detecting the correct orientation is almost
100 percent even for a small number of copies of Pore $\alpha$ (Fig.~\ref{fig:10}(f)). We
would like to stress that it is not necessary to distinguish the orientation of the
polymer before passing them through the pores. Our statistical analysis simply suggest
that it requires far less samples per pore type if we manage to do so.

Our result needs to be compared with the case where heterogeneous segments were
distinguished by their relative interactions with the pore~\cite{Cohen2012}. It turns out
that in such a scenario, the translocation time distributions have sharper and more
distinguished features, leading to better sequencing accuracy. However, as argued before,
structural heterogeneity of a polymer is an experimentally relevant scenario and our
analysis shows that using different patterned pores can lead to efficient sequencing
strategies for such cases. It is important to note that our analysis is robust with
respect to changes in pore width and the length of the polymer (see the supplementary
material).

The sequencing theme outlined above, can be experimentally realized using fabricated
nanofluidic channels with surface decoration. Arrays of nanochannels interfaced with
microfludic loading channels have been shown to be a highly parallel platform for the
restriction mapping of DNA ~\cite{Persson2010,Reisner2010,Schoch2008}.  The first task is
to construct the set of translocation time distributions for known sequences.  This
requires passing sequences with a particular orientation multiple times through these
functionally modified nanofluidic channels. Solid state nanopores are other highly
plausible candidates to achieve the same. With the training set characterised, the
sequencing of heteropolymers with ``unknown'' sequences can be efficiently achieved in
limited time by passing them through these channels using our purely statistical analysis.
The detection of the orientation of the polymer could be achieved using a fluorescent dye
on either the stiff or flexible end of the polymer\cite{Ju1995}.

\section{Summary} \label{sec:summary}

We have shown how statistical fluctuations in the translocation time dynamics could be
efficiently used to sense sequence dependent bending rigidity of biopolymers.  The mean
waiting times, $w(s)$, of the beads of the polymer and correspondingly the mean
translocation times, gives extensive information of the translocation dynamics. The strong
dependence of these properties on the bending rigidity of the polymer and the
distinguishable translocation time statistics generated due to different patterned
stickiness, allows us to efficiently detect polymers with varying bending rigidity by
combining readouts from multiple pores for rapid convergence. For extended pores, the
breakup of the total translocation time into the filling, transfer and escape times proves
useful and in this context reveal interesting features for semiflexible polymer
translocation hitherto unobserved for pores of unit length.  The effect of changing the
external bias and pore length are important aspects of future study.

\section*{Supplementary Material}

See supplementary material for the robustness of our sequence detection scheme with
respect to changes in the pore width and the length of the polymer.

\section*{Acknowledgements}

The authors would like to thank the HPC facility at IISER Mohali for computational time.
The authors thank Debasish Chaudhuri for valuable suggestions and a careful reading of the
manuscript. A.C. acknowledges SERB Project No. EMR/2014/000791 for financial support.
Authors acknowledge Department of Science and Technology, India, for financial support.

\onecolumngrid
\renewcommand{\thefigure}{S\arabic{figure}}
\setcounter{figure}{0}

\newpage

\begin{center}
	{\bf Supplementary Material}
\end{center}

The pore widths used in our study compare favorably with experimental scenarios. The units
of energy, length and mass are set by $\epsilon$, $\sigma$, and $m$, respectively. This
sets the unit of time as $(m \sigma^2 / \epsilon)^{1/2}$ and that of force as
$\epsilon/\sigma$. Following earlier studies ~\cite{Luo2008,Cohen2012}, we assume the bead
size in our coarse-grained polymer model as $\sigma = 1.5$ nm.  This is equal to the Kuhn
length of a single-stranded DNA, which is approximately three nucleotide bases. Hence the
mass of the bead is $m \approx 936$ amu (given that the mass of a base in DNA is $\approx
312$ amu) and charge of the bead $q \approx 0.3$ e (each base having a charge of 0.1 e
effectively). To allow comparison with known results, we set $\zeta = 0.7$ and $k_BT =
1.2$. At $T=295$ K, the interaction strength is given by $\epsilon = k_B T/1.2 \approx 3.4
\times 10^{-21}$ J, which gives a time scale of $(m \sigma^2 / \epsilon)^{1/2} \approx 30$
ps and force scale of $\epsilon/\sigma \approx 2.4$ pN.  Therefore, an external driving
force of $F=1.0$ corresponds to a voltage range $V = FL/q \approx 190-380$ mV across the
pores. In these units, a pore width of $W = 2.25$ corresponds to $\sim 3.5$nm which is not
unphysical.  The internal constriction of the $\alpha$ Hemolysin pore is $\sim 1.5$nm.
Solid state nanopores of width $3-10$nm are now routinely used.

\begin{figure*}[h]

	\centering
  \includegraphics[scale=0.98]{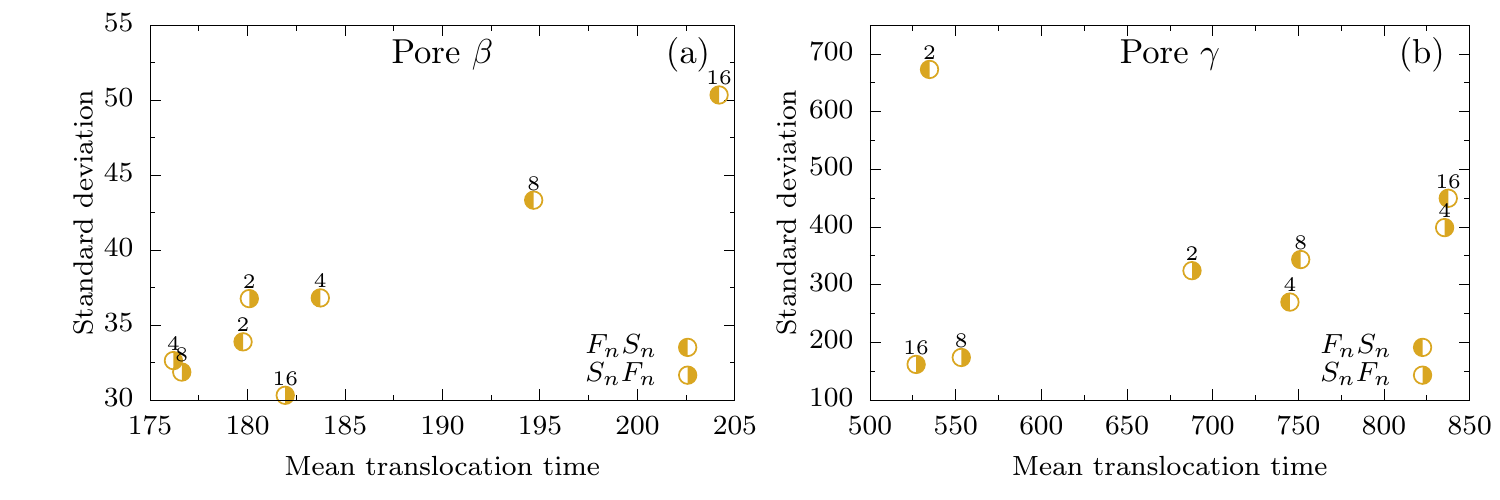}
  \includegraphics[scale=0.98]{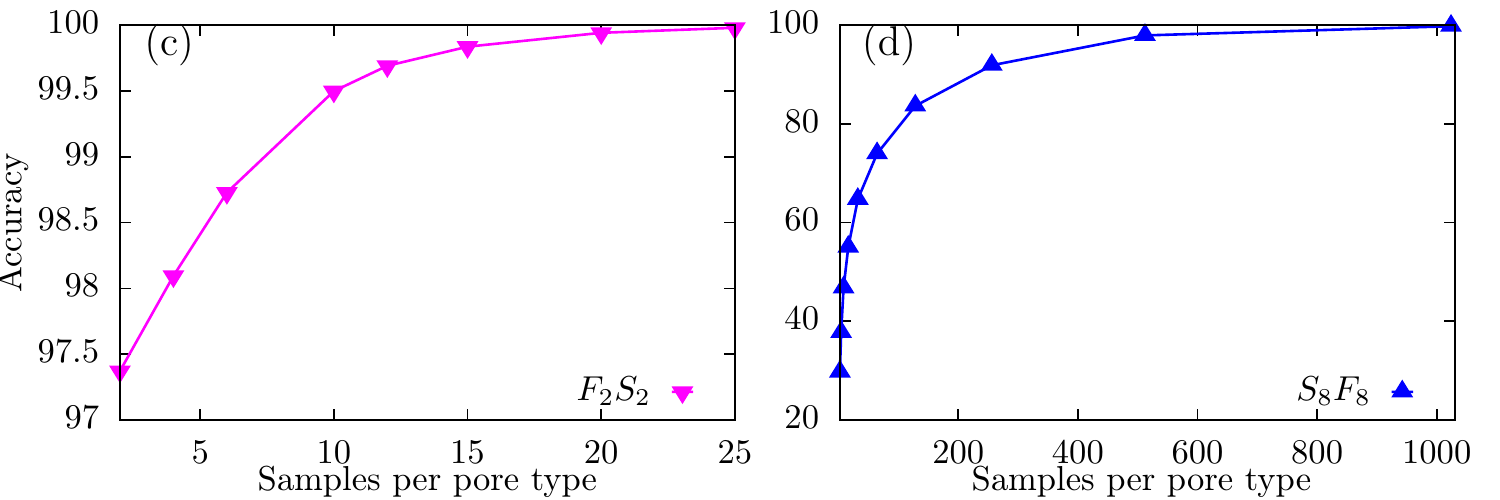}
    
	\caption{\label{R1}(a) and (b) Plot of mean translocation time and standard deviation,
		obtained from 2000 successful translocation events, for a polymer of length $N=32$
		translocating through pore of width $W=3$. The polymer entering the pore from the
		stiff end ($S_nF_n$) is shown by symbol \textcolor{goldenrod}{$\RIGHTcircle$} while
		the polymer entering from the flexible end ($F_nS_n$) is shown by
		\textcolor{goldenrod}{$\LEFTcircle$}. The accuracy of detecting heteropolymers (c)
		$F_{2}S_{2}$ (\textcolor{magenta}{$\blacktriangledown$}) and (d) $S_{8}F_{8}$
	(\textcolor{blue}{$\blacktriangle$}) through Pores $\beta$ and $\gamma$. } 

\end{figure*}

To test the robustness of the method proposed in the main paper, we apply it to
heteropolymer chains consisting of stiff and flexible segments ($S_nF_n$ and $F_nS_n$) of
length $N=32$ translocating through patterned Pores $\beta$ and $\gamma$ having pore width
$W=3$. We choose $\epsilon_{\textrm{pore}} = 2$ for Pore $\beta$ and
$\epsilon_{\textrm{pore}} = 3$ for $\gamma$. All other simulation parameter values are the
same as in the main paper. We construct scatter plot by calculating the mean translocation
time and standard deviation obtained from 2000 successful translocation events. These
plots are shown in Fig.~\ref{R1}(a) and ~\ref{R1}(b) for Pores $\beta$ and $\gamma$,
respectively.  Figures~\ref{R1}(c) and (d) show the accuracy of detection for sequences
$F_{2}S_{2}$ and $S_{8}F_{8}$, respectively. For sequence $F_{2}S_{2}$, the accuracy
reaches $100\%$, even for $~\sim 25$ samples per pore type. This shows that our analysis
is robust enough with respect to changes in the pore width, and the polymer length and an
unknown sequence could be detected to a high accuracy with a relatively small number of
samples per pore type.

\begin{thebibliography}{10}


\bibitem{Frank2010} 
J. Frank and R. L. Gonzalez Jr., Ann. Rev. Biochem. {\bf 79}, 381 (2010).

\bibitem{Salman2001} 
H. Salman et. al., Proc. Natl. Acad. Sci. USA {\bf 98}, 7247 (2001).

\bibitem{Service2006}
R. F. Service, 
Science {\bf 311}, 1544 (2006).

\bibitem{Lagerqvist2006}
J. Lagerqvist, M. Zwolak \& M. Di Ventra,
Nano Lett. {\bf 6}, 779 (2006).

\bibitem{Branton2008}
D. Branton et al. 
Nature Biotechnol. {\bf 26}, 1146 (2008).

\bibitem{Shendure2008}
J. Shendure and H. Ji, 
Nature Biotechnol. {\bf 26}, 1135 (2008).

\bibitem{Schloss2008}
J. A. Schloss,
Natute Biotechnol. {\bf 26}, 1113 (2008).

\bibitem{Persson2010}
F. Persson and J. O. Tegenfeldt,
Chem. Soc. Rev. {\bf 39}, 985 (2010). 

\bibitem{Zwolak2008}
M. Zwolak and  M. Di Ventra,
Rev. Mod. Phys. {\bf 80}, 141 (2008).

\bibitem{Min2011}
S. K. Min, W. Y. Kim, Y. Cho and K. S. Kim,
Nature Nanotechnol. {\bf 6}, 162 (2011).

\bibitem{Deamer2002}
D. W. Deamer and D. Branton,
Acc. Chem. Res. {\bf 35}, 817 (2002).

\bibitem{Kasianowicz1996}
J. J. Kasianowicz, E. Brandin, D. Branton and D. W.  Deamer,
Proc. Natl. Acad. Sci. U.S.A. {\bf 93}, 13770 (1996).

\bibitem{Braha1997}
O. Braha et al.,
Chem. Biol. {\bf 4}, 497 (1997).

\bibitem{Akeson1999}
M. Akeson, D. Branton, J. J. Kasianowicz, E. Brandin and D. W. Deamer,
Biophys. J. {\bf 77}, 3227 (1999).

\bibitem{Meller2000}
A. Meller, L. Nivon, E. Brandin, J. Golovchenko and D. Branton,
Proc. Natl. Acad. Sci. USA {\bf 97}, 1079 (2000).

\bibitem{Meller2001}
A. Meller, L. Nivon and D. Branton,
Phys. Rev. Lett. {\bf 86}, 3435 (2001).

\bibitem{Meller2003}
A. Meller, 
J. Phys. Condens. Matter {\bf 15}, R581 (2003).

\bibitem{Maglia2008}
G. Maglia, M. R. Rincon Restrepo, E. Mikhailova, and H. Bayley, Proc. Natl. Acad. Sci. USA
{\bf 105}, 19720, (2008).

\bibitem{Muthukumar1999}
	M. Muthukumar, J. Chem. Phys. {\bf 111}, 10371 (1999).

\bibitem{lubensky}
D. K. Lubensky and D. R. Nelson, 
Biophys. J. {\bf 77}, 1824 (1999).

\bibitem{luo2006jcp}
I. Huopaniemi, K. Luo, T. Ala-Nissila and S. C. Ying,
J. Chem. Phys. {\bf 125}, 124901 (2006).

\bibitem{muthukumar}
M. Muthukumar and C. Y. Kong,
Proc. Natl. Acad. Sci. USA {\bf 103}, 5273 (2006).

\bibitem{Polson2013}
J. M. Polson and A. C. M. McCaffrey, J. Chem. Phys. {\bf 138}, 174902 (2013).

\bibitem{matysiak}
S. Matysiak, A. Montesi, M. Pasquali, A. B. Kolomeisky and C. Clementi,
Phys. Rev. Lett. 96, 118103 (2006).

\bibitem{luo2007prl}
K. Luo, T. Ala-Nissila, S. C. Ying and A. Bhattacharya,
Phys. Rev. Lett. {\bf 99}, 148102 (2007).

\bibitem{luo2008pre}
K. Luo, T. Ala-Nissila, S. C. Ying and A. Bhattacharya,
Phys. Rev. E {\bf 78}, 061918 (2008).

\bibitem{luo2008prl}
K. Luo, T. Ala-Nissila, S. C. Ying and A. Bhattacharya,
Phys. Rev. Lett. {\bf 100}, 058101 (2008).

\bibitem{luo2007jcp}
K. Luo, T. Ala-Nissila, S-Chen Ying and Aniket Bhattacharya
J. Chem. Phys. {\bf 126}, 145101 (2007).

\bibitem{mirigian2012} 
S. Mirigian, Y. Wang and M. Muthukumar
J. Chem. Phys. {\bf 137}, 064904 (2012).

\bibitem{gauthier}
M. G. Gauthier and G. W. Slater,
J. Chem. Phys. {\bf 128}, 175103 (2008).

\bibitem{luan}
B. Luan et. al.,
Phys. Rev. Lett. {\bf 104}, 238103 (2010).

\bibitem{nikoubashman}
A. Nikoubashman and C. N. Likos,
J. Chem. Phys. {\bf 133}, 074901 (2010).

\bibitem{sung}
W. Sung and P. J. Park,
Phys. Rev. Lett. {\bf 77}, 783 (1996).

\bibitem{muthukumar2001}
M. Muthukumar,
Phys. Rev. Lett. {\bf 86}, 3188 (2001).

\bibitem{kantor2001}
J. Chuang, Y. Kantor, Y. and M. Kardar,
Phys. Rev. E {\bf 65}, 011802 (2001).

\bibitem{muthukumar2003}
M. Muthukumar,
J. Chem. Phys. {\bf 118}, 5174 (2003).

\bibitem{metzler}
R. Metzler and J. Klafter,
Biophys. J. {\bf 85}, 2776 (2003).

\bibitem{slonkina}
E. Slonkina and A. B. Kolomeisky,
J. Chem. Phys. {\bf 118}, 7112 (2003).

\bibitem{kantor2004}
Y. Kantor and M. Kardar,
Phys. Rev. E {\bf 69}, 021806 (2004).

\bibitem{milchev2004}
A. Milchev, K. Binder and A. Bhattacharya,
J. Chem. Phys. {\bf 121}, 6042 (2004).

\bibitem{gerland}
U. Gerland, R. Bundschuh and T. Hwa,
Phys. Biol. {\bf 1}, 19 (2004).

\bibitem{gopinathan}
A. Gopinathan and Y. W. Kim,
Phys. Rev. Lett. {\bf 99}, 228106 (2007).

\bibitem{wong}
C. T. A. Wong and M. Muthukumar,
J. Chem. Phys. {\bf 133}, 045101 (2010).

\bibitem{milchev2011}
A. Milchev,
J. Phys. Condens. Matter {\bf 23}, 103101 (2011).

\bibitem{abdolvahab2011}
R. H. Abdolvahab, M. R. Ejtehadi, and R. Metzler, 
Phys. Rev. E {\bf 83}, 011902 (2011).

\bibitem{Cohen2011}
J. A. Cohen, A. Chaudhuri, and R. Golestanian
Phys. Rev. Lett. {\bf 107}, 238102 (2011). 

\bibitem{Sakaue2007}
T. Sakaue, Phys. Rev. E {\bf 76}, 021803 (2007).

\bibitem{Sakaue2010}
T. Sakaue, Phys. Rev. E {\bf 81}, 041808 (2010).

\bibitem{Saito2011}
T. Saito and T. Sakaue, Eur. Phys. J. E {\bf 34}, 135 (2011).

\bibitem{Rowghanian2011}
P. Rowghanian and A. Y. Grosberg, J. Phys. Chem. B {\bf 115}, 14127 (2011).

\bibitem{Dubbeldam2012}
J. L. A. Dubbeldam, V. G. Rostiashvili, A. Milchev, and
T. A. Vilgis, Phys. Rev. E {\bf 85}, 041801 (2012).

\bibitem{Ikonen2012}
T. Ikonen, A. Bhattacharya, T. Ala-Nissila and W. Sung,
Phys. Rev. E {\bf 85}, 051803 (2012).

\bibitem{Ikonenjcp2012}
T. Ikonen, A. Bhattacharya, T. Ala-Nissila and W. Sung,
J. Chem. Phys. {\bf 137}, 085101 (2012).

\bibitem{Sarabadani2014}
	J. Sarabadani, T. Ikonen and T. Ala-Nissila, 
	J. Chem. Phys. {\bf 141}, 214907 (2014).

\bibitem{Sarabadani2017}
	J. Sarabadani, T. Ikonen, H.M\"{o}kk\"{o}nen, T Ala-Nissila, S. Carson and M. Wanunu,
	Scientific Rep. {\bf 7}, 7423 (2017).        


\bibitem{Lehtola2009}
V. V. Lehtola, R. P. Linna, and K. Kaski, EPL {\bf 85}, 58006 (2009).


\bibitem{Haan2010}
H. W. de Haan, and G. W. Slater, Phys. Rev. Lett. {\bf 110}, 048101 (2013).

\bibitem{Bhattacharya2013}
A. Bhattacharya, Polymer Science, Ser. C {\bf 55},  60  (2013).

\bibitem{Adhikari2013}
R. Adhikari and A. Bhattacharya, J. Chem. Phys., {\bf 138},  204909  (2013).

\bibitem{Adhikari2015}
R. Adhikari and A. Bhattacharya, Europhys. Lett., {\bf 109},  38001  (2015).

\bibitem{Cohen2012}
J.~A. Cohen, A. Chaudhuri, and R. Golestanian, Phys. Rev. X, {\bf 2},  021002  (2012). 

\bibitem{Cohen2013}
J.~A. Cohen, A. Chaudhuri, and R. Golestanian, J. Chem. Phys. {\bf 137}, 204911 (2012).

\bibitem{Katkar2014}
H. H. Katkar and M. Muthukumar, J. Chem. Phys. {\bf 140}, 135102 (2014).

\bibitem{Dhar2002}
	A. Dhar and D. Chaudhuri, Phys. Rev. Lett. {\bf 89}, 065502 (2002).

\bibitem{Frey1996}
	J. Wilhelm and E. Frey, Phys. Rev. Lett. {\bf 77}, 2581 (1996).

\bibitem{Widom2001}
J. Widom, Q Rev Biophys {\bf 34}, 269 (2001).

\bibitem{Travers2004}
A. A. Travers, Philos Trans Roy Soc A {\bf 362}, 1423 (2004).

\bibitem{Geggier2010}
S. Geggier and A. Vologodskii, Proc. Natl. Acad. Sci USA {\bf 107}, 15421 (2010).

\bibitem{Branden1998}
C. Branden and J. Tooze, Introduction of Protein Structure (Garland Publishing, New York, 1998).

\bibitem{Lee2001}
M. Lee, B. K. Cho, and W. C Zin, Chem. Rev. 101, 3869 (2001).

\bibitem{Stellwagen2003}
E. Stellwagen, Y. Lu, and N. Stellwagen, Biochemistry 42, 11 745 (2003).

\bibitem{Kowalczyk2010}
S. W. Kowalczyk, A. R. Hall, and C. Dekker, Nano Lett. 10, 324 (2010).

\bibitem{Storm2003}
A. J. Storm, J. H. Chen, X. S. Ling, H. W. Zandbergen and C. Dekker,
Nature Mater. {\bf 2}, 537 (2003).

\bibitem{Kim2006}
M. J. Kim, M. Wanunu, D. C. Bell and A. Meller,
Adv. Mater. {\bf 18}, 3149 (2006).

\bibitem{Ohshiro2006}
T. Ohshiro and Y. Umezawa, 
Proc. Natl. Acad. Sci. USA {\bf 103}, 1014 (2006).

\bibitem{Iqbal2007}
S. M. Iqbal, D. Akin and R. Bashir, 
Nature Nanotechnol. {\bf 2}, 243 (2007).

\bibitem{Wanunu2007}
M. Wanunu and A. Meller, 
Nano Lett. {\bf 7}, 1580 (2007).

\bibitem{Chen2004}
P. Chen et al., 
Nano Lett. {\bf 4}, 1333 (2004).

\bibitem{Tabard2007}
V. Tabard-Cossa, D. Trivedi, M. Wiggin, N. N. Jetha and A. Marziali,
Nanotechnology {\bf 18}, 305505 (2007).

\bibitem{Reisner2010}
W. Reisner et al.,
Proc. Natl. Acad. Sci. USA {\bf 107}, 13294 (2010).

\bibitem{Schoch2008}
R. B. Schoch, J. Han and P. Renaud,
Rev. Mod. Phys. {\bf 80}, 839 (2008).

\bibitem{Ju1995}
J. Ju, C. Ruan, C. W. Fuller, A. N. Glazer and R. A. Mathies,
Proc. Natl. Acad. Sci. USA {\bf 92}, 4347 (1995).


\end{thebibliography}

\begin{thebibliography}{10}

\bibitem{Luo2008}
K. Luo, T. Ala-Nissila, S. C. Ying and A. Bhattacharya,
Phys. Rev. Lett. {\bf 100}, 058101 (2008).

\bibitem{Cohen2012}
J.~A. Cohen, A. Chaudhuri, and R. Golestanian, Phys. Rev. X, {\bf 2},  021002  (2012).

\end{thebibliography}
\end{document}